\documentclass[aps,prl,twocolumn,showpacs,preprintnumbers,amsmath,amssymb]{revtex4-1}
\usepackage[utf8]{inputenc}
\usepackage[T2A]{fontenc}

\usepackage{upgreek}
\usepackage{graphicx} % Include figure files
\usepackage{bm} % bold math
\usepackage{MnSymbol}
\usepackage[dvipsnames]{xcolor}

\usepackage{color}

\begin{document}

\title{Thermodynamic Evidence for Density Wave Order in Two Dimensional $^{4}$He Supersolid}
\author{J. Knapp, J.Ny\'{e}ki, H. Patel, F Ziouzia, B.P.Cowan and J.Saunders}
\affiliation{Department of Physics, Royal
Holloway University of London, Egham, Surrey, TW20 0EX, U.K}
\date{\today}

\date{\today}

\begin{abstract}
We previously reported the discovery of a two dimensional $^{4}$He supersolid; a state with intertwined density wave and superfluid order, observed in the second layer of $^{4}$He adsorbed
on graphite.
In this Letter we provide direct evidence for the density wave order, obtained by doping the layer with a small concentration of $^{3}$He atoms (impuritons).
The heat capacity, magnetization and NMR relaxation times of the $^{3}$He were measured over a wide temperature range from 200 $\mu$K to 500 mK.
They provide evidence for changes in the ground state in the second layer film as the amount of $^{4}$He is increased at various fixed $^{3}$He doses.
Clear evidence is obtained for a solid second layer film, matching the r\a'egime of superfluid response previously observed in pure $^{4}$He films.
\end{abstract}
\maketitle

%%%%%%%%%%%%%%%%%%%%%%%%%%%%%%%%%%%%%%%%%%%%%%%%%%%%%%%%

% INTRO
Quantum zero point energy plays a crucial role in the ground state properties of helium crystals, due to combination of small atomic mass and the weak interatomic van de Waals potential~\cite{deBoer1948,Simon1934}.
$^{4}$He is a liquid at absolute zero, and when it solidifies under a pressure of 25~bar its molar volume is significantly larger, by almost a factor of three, than expected classically from its potential energy~\cite{Dugdale1953}, an effect more pronounced for $^{3}$He.
The large zero point motion of atoms about equilibrium~\cite{Andreev1982} leads to them moving between sites; this quantum exchange is now understood to occur via atomic ring permutations~\cite{Roger1983}.

Two-dimensional crystalline phases of helium in atomically layered films, grown on the atomically flat surface of graphite, provide the opportunity to study more extreme quantum solids in which quantum exchange is enhanced both by relatively low density and by atomic zero point motion normal to the substrate surface~\cite{Saunders2020}.
As such, this system is a prime candidate to observe the supersolid phase of matter.
In its purest sense, a supersolid is a homogeneous phase of matter in which both superfluid and crystalline order coexist~\cite{Boninsegni2012}.
It has been proposed that the ground state of any Bose solid is supersolid~\cite{Anderson2009,Anderson2012}.

% SUPERSOLID
In pure $^{4}$He quantum crystals a supersolid phase has been proposed~\cite{Leggett1970,Chester1970,Andreev1969}, see~\cite{Boninsegni2012} for a review.
In one scenario~\cite{Chester1970,Andreev1969} supersolidity relies on crystalline incommensurability, arising from spontaneous zero point vacancies free to tunnel through the crystal.
Alternatively, frozen-in crystalline defects such as dislocations~\cite{Boninsegni2007,Soyler2009}, or grain boundaries~\cite{Pollet2007}, can host superfluidity.
The search for supersolidity in bulk solid $^{4}$He has been the subject of intense experimental investigation, in which the smallness of the superfluid fraction together with the inevitable visco-elastic response of the solid have prevented unambiguous detection~\cite{Chan2013}.
A possible exception is the supersolid response of dislocation cores~\cite{Hallock2008,Hallock2010,Hallock2012,Hallock2014}. Recent work using a carbon nanotube as a substrate has also identified possible pathways to a 2D helium supersolid \cite{Todoshchenko2022,Todoshchenko2024}.

Evidence has been found for transient supersolid behavior in quantum gases with long range dipolar interactions, which can be tuned to create an array of phase coherent superfluid droplets~\cite{Tanzi,Chomaz,Bottcher,Tanzi2019,Guo2021,Norcia2021,Tanzi2021}.
Several other approaches have also attempted to engineer cold-atom supersolid light-mediated interactions~\cite{Leonard} or spin orbit coupling~\cite{Li}.

\begin{figure}
    \includegraphics{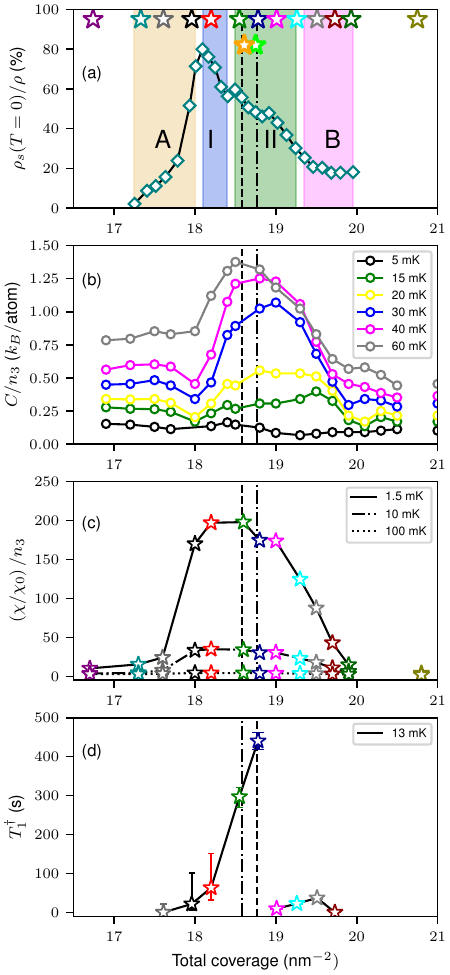}
    \caption{
    Isotherms as a function of total coverage:
    (a) torsional oscillator superfluid fraction at $T=0$ (pure $^{4}$He \cite{Nyeki2017,Nyeki2017b});
    (b) heat capacity ($^{3}$He coverage fixed at 0.7 nm$^{-2}$);
    (c) Normalised $^{3}$He nuclear magnetic susceptibility ($^{3}$He coverage fixed at 0.51 nm$^{-2}$ below 18.8 nm$^{-2}$, and 0.74 nm$^{-2}$ above);
    (d) NMR effective spin--lattice relaxation time $T_1^\dagger$, as a signature of 7/12 commensuration.
    Vertical lines mark position of the 4/7 and 7/12 commensurate coverages.
    }
    \label{fig:Isotherms}
\end{figure}

% RHUL TO EXPERIMENT
In a previous report, we have discussed evidence for an emergent two-dimensional supersolid in the second layer of $^{4}$He on graphite~\cite{Nyeki2017,Nyeki2017b}.
That torsional oscillator study of the mass decoupling of the film was made over a fine grid of coverages to significantly lower temperatures than prior work~\cite{Crowell1996}, approaching 1\,mK.
Achieving such low temperatures was essential to establish both the temperature dependence of the superfluid density and the full coverage range over which the supersolid response occurs.
A superfluid fraction at $T=0$ ranging from 0.8 to 0.2 was established.
Furthermore, a remarkable scaling behavior of the superfluid density was identified.
We found a sequence of four coverage intervals with distinct features of data collapse; two with scaling of a single parameter which is a characteristic temperature scale (I, II) and two with two-parameter scaling (A, B), Fig.~\ref{fig:Isotherms}\,(a).
This provided further evidence of the potential interplay between film structure and superfluid response.
In the second layer supersolid, the anomalous temperature dependence of the superfluid density in the low temperature limit was explained in terms of a spectrum of elementary excitations with a set of soft roton minima.
It follows that the structure factor is strongly peaked at the momenta of these minima: density wave order.
These results led us to propose a state of intertwined density wave and superfluid order.
Supportive evidence for superfluidity, as opposed to visco-elastic response, is provided by a torsional oscillator study which shows that the superfluid density is independent of frequency~\cite{Choi2021}.
However, because the minimum temperature in that work was 20\,mK it was not possible to follow the superfluid response over the full second layer coverage range.

% SIMULATIONS
Importantly, the recent simulations by quantum Monte Carlo methods~\cite{Gordillo2020} find evidence of supersolidity (the coverage scale is within 1\% of that adopted in our work).
A stable 7/12 second layer superlattice phase is identified, with superfluid fraction 0.3.
This r\a'egime of supersolid response is well aligned with regions I and II, identified experimentally.
This recent theoretical result follows a series of earlier predictions investigating the stability of a crystalline second-layer phase of $^{4}$He on both graphite and graphene, by a variety of first principles simulation methods, at variance with experimental observations.
While there is consensus on the stability of an incommensurate second layer solid phase at around 7.6\,nm$^{-2}$, earlier simulations~\cite{Corboz2008,Gordillo2012,Ahn,Kwon,Happacher} found no solid phase at lower densities at which we find supersolid response and where heat capacity anomalies at 1-1.5\,K provide signatures of melting~\cite{Greywall1991,Greywall1993,Nakamura2016}.
In the r\a'egime of interest, the putative epitaxial triangular superlattices at relative densities of 4/7 and 7/12 have been simulated, but found to be unstable in~\cite{Boninsegni2019,Boninsegni2020,Boninsegni2023}.
These simulations contradict earlier work, which identified a 4/7 phase stable against particle/hole doping over a density range of around 3$\%$~\cite{Pierce1998,Pierce1999}.
By contrast, in the case of $^{3}$He, a stable incommensurate crystalline phase is found at these densities, but with no evidence of pinning of the superlattice structure to the underlying first layer potential~\cite{Gordillo2016,Gordillo2018}, as found experimentally in~\footnote{F. Arnold, Ph.D. thesis (Royal Holloway 2015)}.

The objective of the experiments reported in this Letter was to establish the coverage r\a'egime of solid order in the second layer at ultra-low temperatures, to support the already compelling, but indirect, evidence from torsional oscillator measurements, to complement the signatures of melting at 1-1.5\,K, and to test the theoretical predictions discussed above.
We note that the first experiments which detected an anomalous superfluid response in the second layer~\cite{Crowell1996} suggested, on the basis of the then current second layer phase diagram~\cite{Greywall1991} and alignment of coverage scales, that superfluidity was ``destroyed by solidification of the film''.
We studied the thermodynamics and dynamics of $^{3}$He impurities doped into the second layer at relatively low concentrations, and tuned the $^{4}$He coverage through the region of interest to probe the quantum state of the layer, via the degree of quantum degeneracy \textit{vs.} localization of the $^{3}$He.
Measurements of the $^{3}$He magnetic susceptibility by NMR are particularly illuminating.
The motivation is that in the simplest case, fully localized $^{3}$He impurities in solid $^{4}$He would exhibit Curie law behavior.
On the other hand, mobile $^{3}$He, either within or on top of a liquid $^{4}$He layer, would exhibit a cross-over from Curie law to a temperature independent Pauli susceptibility on cooling through the Fermi temperature, due to the onset of quantum degeneracy.
However, the fact that two dimensional helium is a quantum crystal has important consequences that affects this picture.
It is well established that $^{3}$He impurities in bulk solid $^{4}$He can form, at low concentrations, a gas of delocalized quasi-particles.
These ``impuriton'' excitations quantum mechanically tunnel through the host crystal~\cite{Andreev1969,Andreev1982,Richards,Grigoriev1973}, but are subject to strain-mediated interactions between them.
Importantly, in the putative 2D $^{4}$He quantum crystal the larger tunneling bandwidth and weaker strain interactions have the consequence that the 2D $^{3}$He impuriton gas can enter the quantum degenerate r\a'egime before localization effects play a role at lower temperatures, see Supplementary material~\cite{SM}.
Such effects were previously observed in the first layer~\cite{Saunders1992}.

Despite these considerations, our measurements of the $^{3}$He magnetic susceptibility nevertheless provide a clear signature of the state of the $^{4}$He layer.
In addition, further dynamical information from the NMR relaxation times unexpectedly provides a clear signature of tuning through the 7/12 structure.
We also report measurements of the heat capacity, which is dominated by the contribution from the $^{3}$He impurities.
Since the three experiments (NMR, heat capacity, and torsional oscillator) are performed in different experimental cells, we adopt a simple but robust methodology to establish a consistent coverage scale.
In each case, we identify point-B of a $^{4}$He vapor pressure isotherm at 4.2\,K~\cite{Rapp1993}, and $\it{define}$ it to correspond to the coverage 11.4\,nm$^{-2}$; thus promotion to the second layer provides the fiducial point.
We note that adoption of this straightforward procedure would also facilitate comparison of results from different groups.
Each experiment also gives a signature, from the coverage dependence of the measured quantity, to identify $^{4}$He promotion to the third layer, which we consistently find at 20.0\,nm$^{-2}$ on the adopted coverage scale.
These coverages of promotion to the second and third layer agree with theoretical simulations.

% NMR
The $^{3}$He magnetization and NMR relaxation times were measured in a cell containing exfoliated graphite with surface area 12\,m$^{2}$.
The NMR spectral lines are rather narrow, with $T_{2}^*$ characterizing the free induction decay (FID) varying from 0.5 to 3\,ms, where the upper limit is imposed by inhomogeneities of the static NMR field.
Pulsed NMR was performed using a broadband SQUID NMR spectrometer, typically operating at $^{3}$He Larmor frequency 100\,kHz~\cite{Korber2003,Arnold2014}.
This technique allows us to obtain high precision data over a wide temperature range from 500\,mK down to 200\,$\mu$K.
The sample chamber used for heat capacity measurements was the same as in previous work~\cite{Siqueira1997}, with substrate surface area 182\,m$^{2}$.

A clear visualization of the change of state of the film with increasing $^{4}$He coverage, and its correlation with superfluid response, is provided by comparing isotherms of heat capacity and NMR data with the superfluid fraction of the putative supersolid in Fig.~\ref{fig:Isotherms}.
These isotherms of four distinct physical quantities, taken on three different cells, demonstrate a high degree of alignment.
As discussed in detail below, they support the stability of a solid phase extending from significantly below the density of both the potential 4/7 superlattice (6.8\,nm$^{-2}$, total coverage 18.7\,nm$^{-2}$) and 7/12 superlattice (6.9\,nm$^{-2}$, total coverage 18.8\,nm$^{-2}$), through to promotion to the third layer.
Here we take the density of the compressed first layer to be 11.9\,nm$^{-2}$, on our coverage scale, consistent with neutron scattering data~\cite{Roger1998_b}.

\begin{figure}
    \centering
    \includegraphics{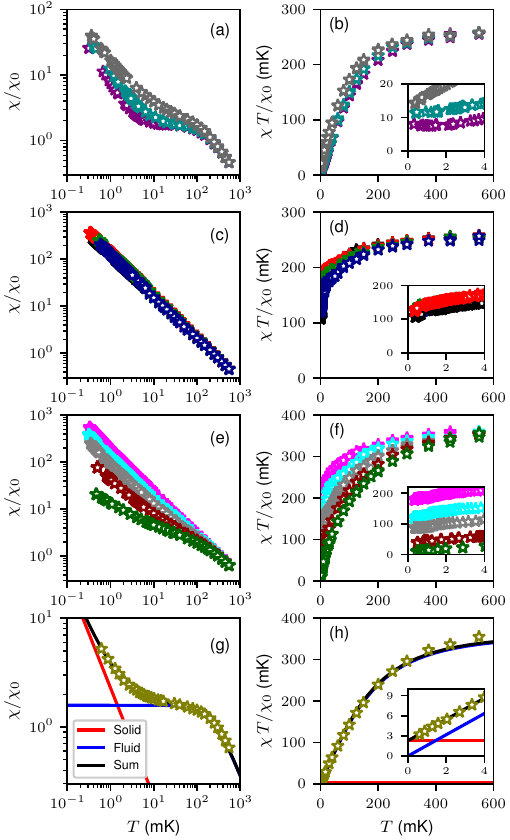}
    \caption{
    Normalized nuclear magnetic susceptibility of $^{3}$He with coverage increasing from (a) to (g).
    The total coverage indicated by the position of stars of equal color in Fig.~\ref{fig:Isotherms}.
    In sub-figures (a,b,c,d) $^3$He coverage was 0.51\,nm$^{-2}$, in (e,f,g,h) 0.74\,nm$^{-2}$;
    (b,d,f,h) use modified coordinates $\chi T/\chi_0$ \textit{vs.} $T$ to emphasize deflection from Curie law.
    Insets to sub-figures (b,d,f,h) highlight $T=0$ intercept indicating a localized fraction;
    (g,h) show a sample in which all $^3$He is promoted into the third layer, behaving mostly as a liquid with sub 1\% solid contribution.
    }
    \label{fig:Susceptibility}
\end{figure}

We also observe an unexpected sharp anomaly in the effective spin-lattice relaxation time on tuning the total coverage.
This feature seems to be a striking signature of the putative 7/12 superlattice.
It signifies that $^{3}$He diffusion is strongly suppressed at the commensurate coverage.
We believe the anomaly arises from the interplay of the in-plane potential, experienced by the $^{3}$He due to the first $^{4}$He solid layer when the second layer is at the commensurate density, with the $^{3}$He-$^{3}$He strain interactions.
Detailed discussion of the spin-lattice relaxation and its interpretation is beyond the scope of this Letter.
In the present context, the observation of the anomaly confirms the theoretical predictions of the stability of the 7/12 state~\cite{Gordillo2020}.
There is no evidence for the 4/7 phase.

% COVERAGE DEPENDENCE - MAIN DISCUSSION
We now consider in more detail the temperature dependence of the magnetization, arising from the dopant $^{3}$He, as it evolves with increasing total coverage, Fig.~\ref{fig:Susceptibility}.
We present our susceptibility normalized by the 2D free Fermi gas Pauli susceptibility; $\chi_0 = \hat{C}/T_F^0 = \mu_0\mu^2mA/\pi\hbar$, which in 2D is independent of density, since the Curie constant $\hat{C}=\mu_0\mu^2N/k_B$ and the Fermi temperature $T_F^0 = \pi\hbar^2N/k_BmA$.
$\chi_0$ is determined from an independent calibration of our NMR magnetometer, and scales our data in Fig.~\ref{fig:Susceptibility}.
We also plot $\chi T/\chi_0$ \textit{vs.} $T$, to emphasize the reduction relative to Curie law, as a signature of the effects of quantum degeneracy on cooling.
See~\cite{SM} for more details.

Below the total coverage 17.6\,nm$^{-2}$, in Fig.~\ref{fig:Susceptibility}\,(a,b), the results are consistent with the Pauli susceptibility of mobile $^{3}$He on or in a fluid $^{4}$He second layer film, with a small fraction of less than a few percent, of localized $^{3}$He spins.
This is most clearly seen in Fig.~\ref{fig:Susceptibility}\,(b) where $\chi T/\chi_0$ \textit{vs.} $T$ is linear-in-$T$ with a small finite intercept at $T=0$.
This result is supported by a linear-in-$T$ heat capacity at the total coverage 17.7\,nm$^{-2}$, shown in Fig.~\ref{fig:Fanoula_heat_capacity}.

\begin{figure}
    \centering
    \includegraphics{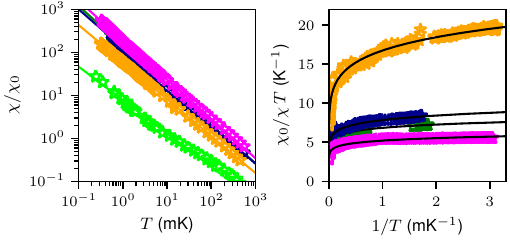}
    \caption{Power law fit of measurements from the vicinity of the commensurate phases over the whole temperature range.
    Samples with varying concentration of $^3$He corresponding to $^3$He coverage 0.1 (light-green), 0.3 (orange), 0.51 (blue and green), and 0.74\,nm$^{-2}$ (magenta).
    See Table~S1 for details and Fig~S6 for the concentration dependence of the fitted exponent. The lowest $^3$He coverage results may be affected by residual substrate heterogeneity.}
    \label{fig:He_power_law_fit}
\end{figure}

With addition of $^{4}$He, in the range 17.96 to 19.01\,nm$^{-2}$, there is a sharp increase of magnetization in the low temperature limit in Fig.~\ref{fig:Susceptibility}\,(c,d) (note different vertical scale of Fig.~\ref{fig:Susceptibility}\,(c) compared to Fig.~\ref{fig:Susceptibility}\,(a)).
This signifies that the state of the host $^{4}$He is now solid.
However, the susceptibility of the $^{3}$He does not follow a simple Curie law as would be expected for full localization of $^{3}$He impurities.
At high temperatures, Fig.~\ref{fig:Susceptibility}\,(d), the initial reduction on cooling from 600\,mK of $\chi T/\chi_0$ below the classical Curie value can be understood in terms of quantum degeneracy arising from the quantum tunneling of $^{3}$He impuritons in a solid $^{4}$He matrix.
Thus, a fit to a Fermi gas determines an effective degeneracy temperature $T_F$, typically 50\,mK, and $^{3}$He band mass, of order 5 times the bare $^{3}$He mass, see~\cite{SM} Fig.~S5.
It is striking that, over this coverage range, the magnetization is extremely well described by a power law dependence, $1/T^\alpha$ with $\alpha\approx0.9$; this extends over the full temperature range from 600\,mK to 200\,{$\mu$}K.
Examples of the fitting are shown in Fig.~\ref{fig:He_power_law_fit} and the dependence of $\alpha$ on $^3$He concentration in Fig.~S6.
It is intriguing that this temperature dependence is similar to that observed in P doped Si on the insulating side of the metal-insulator transition~\cite{Sarachik1986,Bhatt1982,Sachdev1989,Dobrosavljevic1994,Schlager1997,Paalanen1988}.

From 19\,nm$^{-2}$ to 19.93\,nm$^{-2}$, Fig.~\ref{fig:Susceptibility}\,(e,f), the proportion of localized $^{3}$He steadily decreases, as indicated by the zero temperature intercept in Fig.~\ref{fig:Susceptibility}\,(f).
This implies that the impuriton behavior inside the incommensurate solid differs from that in the vicinity of commensuration.
The $^3$He tunneling rate is expected to be influenced by its atomic wave function normal to the surface~\cite{Saunders1992}.
Thus, the observation is consistent with a gradual shift of the $^3$He away from the graphite substrate, until, at the coverage 20.75\,nm$^{-2}$, Fig.~\ref{fig:Susceptibility}\,(g,h), the $^{3}$He is promoted to the third layer, on top of a solid $^{4}$He second layer.
This is indicated by a Pauli susceptibility with a tiny fraction of localized spins (0.61\%).
In summary, the conclusion from the magnetization measurements is that the second layer is solid above around 18\,nm$^{-2}$.
As far as the structure of this solid is concerned, the spin-lattice relaxation time anomaly provides a signature of 7/12 superlattice commensuration.

A consistent picture emerges from the temperature dependence of the heat capacity data, shown for selected coverages in Fig.~\ref{fig:Fanoula_heat_capacity}.
In the vicinity of the density at which the 7/12 superlattice forms, 18.8\,nm$^{-2}$, a maximum in the heat capacity develops, with $\sim$1.3$k_{B}$ per $^{3}$He atom.
At temperatures above the maximum the heat capacity decreases rather slowly with increasing temperature,
while at the lowest temperatures it decreases faster than linearly, as would have been expected for a degenerate gas of $^{3}$He impuritons at $T<<T_{F}$.
This is a further signature of $^{3}$He impuriton interactions.
Above 19.3\,nm$^{-2}$, the evolution of the temperature-dependence of the heat capacity indicates the progressive selective promotion of $^{3}$He to the third layer as seen in the magnetization data.
At 20.33\,nm$^{-2}$, the linear heat capacity confirms that all the $^{3}$He forms a 2D Fermi liquid in the third layer on top of the second layer of solid $^{4}$He.

\begin{figure}[!t]
    \includegraphics{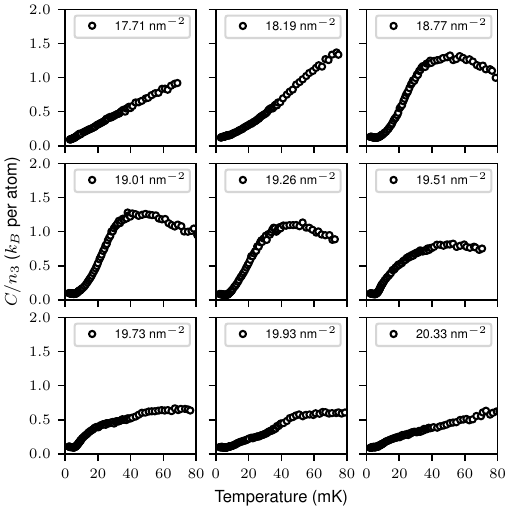}
    \caption{
    Temperature dependence of heat capacity at total coverages shown with $^{3}$He coverage fixed at 0.7\,nm$^{-2}$.
    }
    \label{fig:Fanoula_heat_capacity}
\end{figure}

Finally, we discuss the results in terms of the model of two dimensional $^{3}$He impuritons, for details see~\cite{SM}.
The tunneling rate for a triangular lattice is $t=J_{2}+2J_{3}+4J_{4}+$...~\cite{Saunders1992}, where $J_{n}$ denote atomic ring exchange frequencies of two, three, four, etc. atoms.
Such ring exchange has been confirmed in measurements on pure $^{3}$He films~\cite{Siqueira1997,Roger1990,Siqueira1996,Roger1997,Roger1998,Casey2013}.
We can first make a model of the 2D $^{3}$He impuriton gas~\cite{SM}.
Neglecting interactions between the $^{3}$He impuritons, tunneling leads to the formation of a narrow band Fermi gas, with bandwidth $\Delta = 9t$, bandmass $m_{b }=3\hslash ^{2}/a^{2}\Delta$, and Fermi energy $E_{F}=(2\pi/3\surd 3) \Delta x_{3}$.
This model accounts for the observed characteristic degeneracy temperature and the maximum in the heat capacity data.
The bandwidth of order 0.5\,K is consistent with atomic ring exchange parameters (inferred from $\textit{ab initio}$ calculations on a pure $^{3}$He second layer~\footnote{M. Roger, private communication}). 
Importantly, in 2D the large bandwidth leads to the observed finite solubility of $^{3}$He in 2D $^{4}$He, and the absence of phase separation, which is only seen at much higher $^{3}$He dosage, see~\cite{SM} Fig.~S7.
The same figure however shows the $^{3}$He concentration dependence of the heat capacity maximum does not agree with the simple gas model.
Similarly, we observe a decrease of the bandwidth with increasing $^{3}$He concentration, as determined form the susceptibility measurements, Fig.~S5.
We attribute the breakdown of the model to the $^{3}$He--$^{3}$He impuriton interactions.
At the lowest temperatures, these interactions lead to localization of the $^{3}$He within the solid $^{4}$He, Fig.~\ref{fig:Susceptibility}.
Here theories of random-singlet states in disordered antiferromagnets, which have been applied to P doped Si~\cite{Bhatt1982}, may account for the observed temperature dependence of the magnetization.

In these experiments our goal has been to establish the full coverage regime over which the second layer of $^{4}$He is solid, the stability of which has been theoretically questioned by a number of authors.
Our approach is to introduce small doping levels of $^{3}$He and study their thermodynamic response.
Our working hypothesis is that the $^{3}$He doping does not significantly perturb the structure and stability of the host solid second layer $^{4}$He.
This is reasonable since the transition temperature of the solid order is relatively high~\cite{Greywall1991,Greywall1993,Nakamura2016}.
Here the $^{3}$He dopants are simply being used to assay the structure of the second layer.
The influence of the $^{3}$He dopants on the superfluid response of the pure $^{4}$He supersolid,
onsetting at much lower temperatures, is a quite different and separate question.
The alignment of the isotherms in Fig.~\ref{fig:Isotherms} strongly supports our hypothesis.
The evidence for a solid phase supports our claim that the torsional oscillator response previously observed is that of a supersolid.
We also find evidence for the 7/12 commensuration through a striking signature in the $^{3}$He nuclear spin lattice relaxation time.
Recent theory identifies the 7/12 phase as a supersolid~\cite{Gordillo2020}, with the calculated superfluid fraction 0.3, comparable to our previous measurements~\cite{Nyeki2017,Nyeki2017b}, see Fig.~\ref{fig:Isotherms}.

These results motivate direct measurement of the elastic properties of the pure $^{4}$He film on graphite. Theoretical work on the anomalous elastic properties of a bulk supersolid~\cite{Dorsey2006,Son2005,Heinonen2019,Josserand2007,During2011,Rakic2024} suggest that such measurements of the pure $^{4}$He film should reflect the supersolid order.
To address the technical challenge will likely require nanomechanical studies of films on graphene or carbon nanotubes.

This work was supported in part by EPSRC (UK) through EP/H048375/1, and received funding from the European Union's Horizon 2020 Research and Innovation Programme, under Grant Agreement no. 824109. Measurements were made at the London Low Temperature Laboratory and we thank Richard Elsom, Ian Higgs, Paul Bamford and Harpal Sandhu for technical support. We acknowledge valuable conversations with Michel Roger and Alexander Andreev.

\onecolumngrid

\renewcommand{\thefigure}{S\arabic{figure}}
\renewcommand{\thetable}{S\arabic{table}}
\renewcommand{\theequation}{S\arabic{equation}}
\setcounter{figure}{0}
\setcounter{table}{0}
\setcounter{equation}{0}

\vskip2em

\centerline{\large\textbf{Supplemental Material}}
\vskip1em

\twocolumngrid

\section{Delocalised \textsuperscript{3}He Impuriton Model}

In quantum crystals, where the zero point motion of the atoms is comparable to the lattice parameter, the zero point energy can be lowered by de-localisation~\cite{Boer1957,GUYER1971,Andreev1969,Andreev1982,Richards,Grigoriev1973}.
The \textsuperscript{3}He atom in a \textsuperscript{4}He crystal, is periodically translationally invariant.
Following the nomenclature of prior work on \textsuperscript{3}He atom in a bulk \textsuperscript{4}He crystal we refer to the \textsuperscript{3}He as an \textit{impuriton}~\cite{Andreev1982,Richards,Grigoriev1973}.
In the following, we develop the de-localised quasiparticle thermodynamics of such a system in a 2D layer.

\begin{figure}
    \centering
    \includegraphics[trim=9cm 7cm 2cm 14cm, height=4cm]{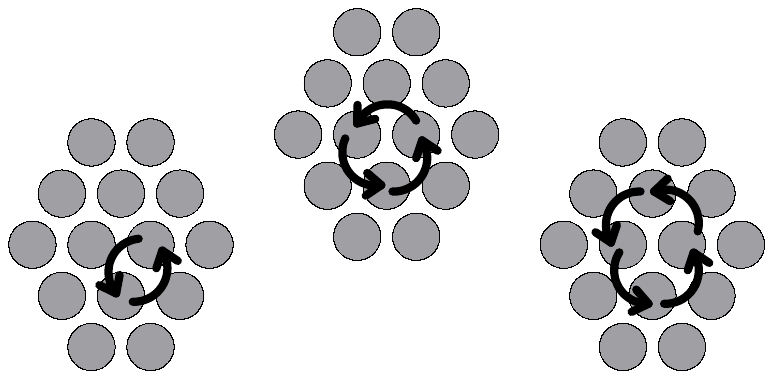}
    \caption{Cyclic exchange of 2, 3 and 4 atoms in a plane.}
    \label{fig:Cyclic_perm}
\end{figure}

In this picture a \textsuperscript{3}He impurity atom is de-localised in the \textsuperscript{4}He matrix, tunneling between sites with a Hamiltonian
\begin{equation}
    \widehat{H} = -t\sum_{i,j} a_i^+a_j,
\end{equation}
where we consider $i$ and $j$ to be nearest neighbors and $t$ is the tunneling amplitude for the nearest neighbor sites.
The tunneling amplitude equals the \textit{sum} of possible cyclic permutations as depicted in Fig.~\ref{fig:Cyclic_perm}, $t=J_2 + 2J_3 + 4J_4 + 2J_6...$, where $J$ is the exchange constant for the appropriate particle exchange.
We make the simplifying assumption, as in 3D mixtures~\cite{Richards1976}, that the tunneling rate is the same for \textsuperscript{3}He--\textsuperscript{4}He and \textsuperscript{4}He--\textsuperscript{4}He pairs.

Thus, while the exchange parameters combine additively to determine the tunneling rate of the \textsuperscript{3}He impuriton, in a pure \textsuperscript{3}He film there is frustration from permutation of odd or even numbers of fermionic atoms.
For example, the effective Heisenberg exchange interaction $J_{\mathrm{eff}}=J_2-2J_3$, so two exchange parameters enter as an inseparable combination.
An estimate of the tunneling rate comes from \textit{ab initio} calculations of exchange interactions in the second layer of solid \textsuperscript{3}He which lead to $t\sim 50$\,mK~\footnote{M. Roger, private communication}.
These large exchange interactions in the second layer lead to a large tunneling bandwidth, which distinguishes the 2D from the 3D case.

In the 3D case it is established that impuritons are subject to long range strain interactions mediated by the solid \textsuperscript{4}He host crystal, of the form $V(r)=V_{0}(a/r)^{3}$, where $a$ is the nearest neighbor distance of the \textsuperscript{4}He triangular lattice.
With increasing impuriton concentration this leads to correlated motion of impuritons and ultimately a localization transition.
This was observed through measurements of the spin diffusion coefficient~\cite{Kagan1992,Kagan1984,Mikheev1983}, and occurs at ${x_{3}\sim 2-5\%}$, depending on molar volume.
The condition for delocalized impurities is that the difference in strain potential between two neighboring sites is less than the bandwidth.
This leads to an upper limit on the \textsuperscript{3}He concentration for the \textsuperscript{3}He atoms to be delocalized
$x\lesssim (\Delta /V_{0})^{3/4}\sim 0.5\%$~\cite{Andreev1982}, where $\Delta$ is the impuriton bandwidth.
In 2D case of the second layer on graphite, the much larger tunneling bandwidth as well as likely weaker interaction energies, allow the study of significantly larger \textsuperscript{3}He concentrations.

Neglecting impuriton interactions, we construct a simple tight--binding model which describes \textsuperscript{3}He quasiparticles within the putative 2D \textsuperscript{4}He matrix.
The following results are for the Fermi energy, bandwidth and impuriton/quasiparticle bandmass
\begin{equation}
    \frac{E_F}{\Delta} = \frac{2\pi}{3\sqrt{3}}x_3 \approx 1.2x_3,
\label{eq:impuritons_Fermi}
\end{equation}
where $x_3$ is the \textsuperscript{3}He concentration, and the effective mass of the band
\begin{equation}
    m_b = \frac{\hbar^2}{3a^2t}.
\label{eq:eff_mass}
\end{equation}
Within this model the bandwidth relates to the tunneling amplitude by
\begin{equation}
    \Delta = 9t.
\end{equation}
The band mass thus depends on the bandwidth as
\begin{equation}
    m_b = \frac{3\hbar^2}{a^2\Delta}.
\label{eq:Band_mass}
\end{equation}

We now discuss the heat capacity and magnetic susceptibility of an ideal gas of fermions in this finite band with this band mass.
The internal energy $U$ of this system is
\begin{equation}
    U = 2\int_0^\Delta \frac{ g(\epsilon)\epsilon }{e^{(\epsilon-\mu)/k_BT}+1} d\epsilon,
\end{equation}
with a constant density of state for $N_3$ \textsuperscript{3}He atoms
\begin{equation}
    g = \frac{N_3}{2E_F},
\end{equation}
it becomes
\begin{equation}
    U = \frac{N_3}{E_F} \int_0^\Delta \frac{ \epsilon }{e^{(\epsilon-\mu)/k_BT}+1} d\epsilon,
\end{equation}
where $\mu$ is the chemical potential.
The chemical potential is found by inverting the formula for number of particles (Fermi--Dirac distribution)
\begin{equation}
    \mu(T) = k_BT\ln\left( \frac{e^{E_F/k_BT} - 1}{1 - e^{(E_F-\Delta)/k_BT}} \right).
\end{equation}

\begin{figure}[!t]
    \centering
    \includegraphics[height=1.6in]{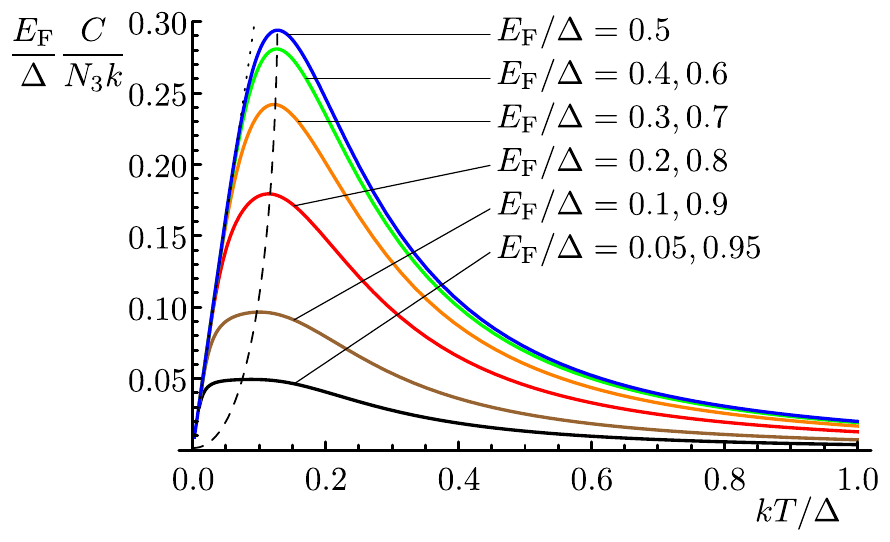}
    \caption{Heat capacity according to the finite bandwidth model, scaled by the ratio of bandwidth and Fermi energy.}
    \label{fig:Finite_band_heat_capacity}
\end{figure}

The numerator is the ideal Fermi gas chemical potential while the denominator accounts for the finite bandwidth.
The heat capacity $C=\partial U/\partial T$ is a complicated expression, but its qualitative description can be given.
At low temperatures the bandwidth $\Delta$ has no influence and the heat capacity is linear-in-$T$
\begin{equation}
    C = \frac{\pi^2k_B^2}{3}\frac{N_3}{E_F}T.
\end{equation}
At higher temperatures, the formula deviates from the linear behavior, with a maximum that arises from the finite bandwidth.
This maximum moves to higher temperatures and increases in magnitude with increasing $E_F/\Delta$ ratio, until $E_F/\Delta = 0.5$.
The behavior is symmetrical about half-filling.
However, we practically only expect the model to work at low concentrations, where interactions are relatively weak.

\begin{figure}[!t]
    \centering
    \includegraphics[height=1.6in]{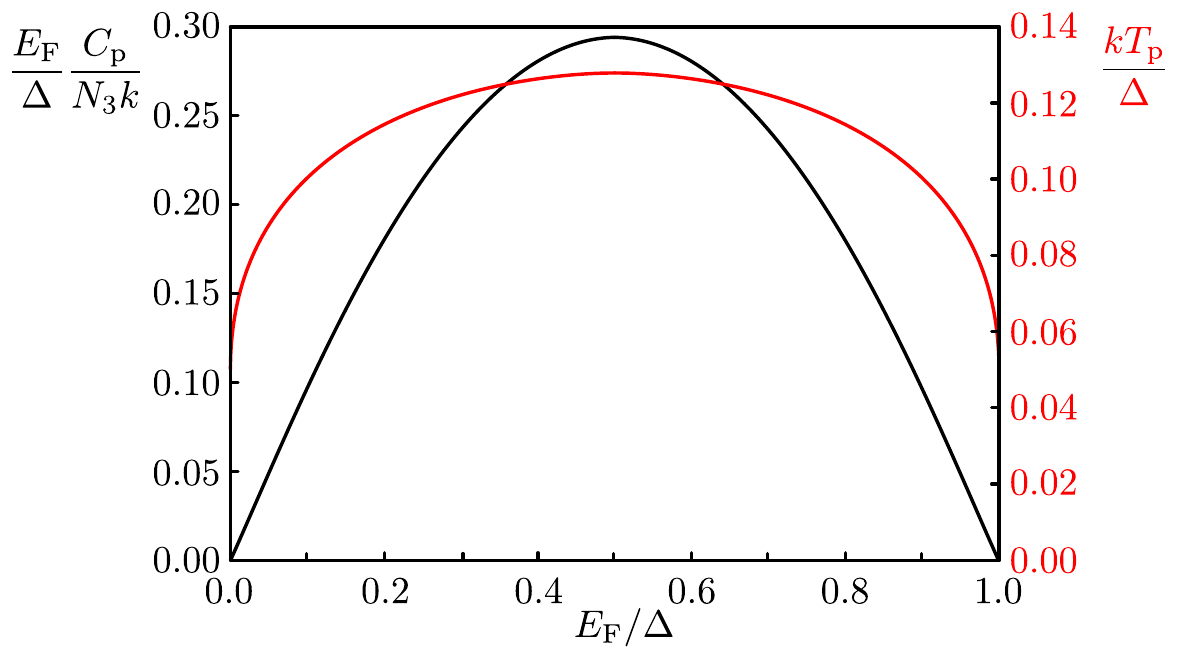}
    \caption{Position and magnitude of the heat capacity maximum according to the finite bandwidth model.}
    \label{fig:Finite_band_model_peak_pos_mag}
\end{figure}

The maximum significantly broadens in the $E_F/\Delta\rightarrow0$ limit.
The high temperature expansion of the heat capacity formula has the leading $T^{-2}$ temperature dependence, characteristic of a Schottky peak
\begin{equation}
    \frac{C}{N_3} = \frac{\Delta(\Delta-E_F)}{12k_BT^2} + \frac{\Delta(-\Delta^3 + 2\Delta^2E_F - 2\Delta E_F^2 + E_F^3)}{240k_B^3T^4} + ...
\end{equation}
The overall temperature dependence of the heat capacity is shown in Fig.~\ref{fig:Finite_band_heat_capacity}.
When scaled by $E_F/\Delta$, curves for ratio $x$ and $1-x$ collapse onto each other.
The position and magnitude of the heat capacity maximum is shown in Fig.~\ref{fig:Finite_band_model_peak_pos_mag}.

\begin{figure}[!t]
    \centering
    \includegraphics[height=1.6in]{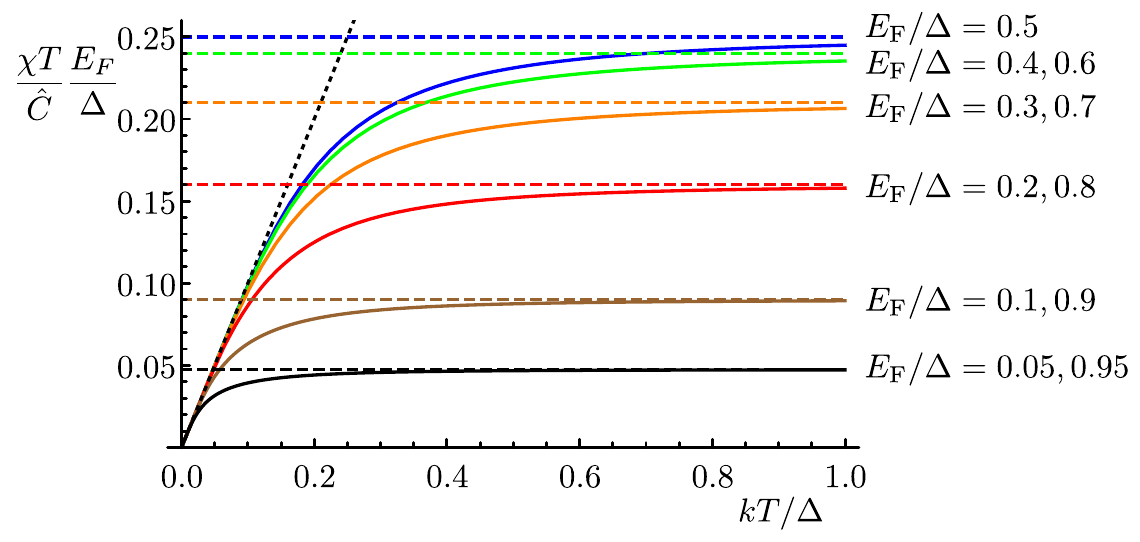}
    \caption{Susceptibility according to the Finite bandwidth model, scaled by the ratio of bandwidth and Fermi energy.}
    \label{fig:Finite_band_susceptibility}
\end{figure}

The magnetic susceptibility is derived from the chemical potential as
\begin{equation}
    \chi = \frac{k_B\hat{C}}{N_3\frac{\partial\mu}{\partial N_3}},
\end{equation}
where the Curie constant is
\begin{equation}
    \hat{C} = \frac{\mu_0\mu^2N_3}{k_B}.
\label{eq:Curie_constant_FB}
\end{equation}
We derive
\begin{equation}
    \chi(T) = \frac{2k_B\hat{C}}{E_F} \frac{\sinh(E_F/2k_BT)\sinh((\Delta-E_F)/2k_BT)}{\sinh(\Delta/2k_BT)},
\label{eq:finite_band_susceptibility}
\end{equation}
shown in Fig.~\ref{fig:Finite_band_susceptibility} in $\chi T$ \textit{vs.} $T$ coordinates.
When scaled by the $E_F/\Delta$ ratio the particle--hole symmetry is apparent from the graph, similar to the heat capacity.
High temperature expansion of equation~\ref{eq:finite_band_susceptibility} recovers Curie law $\chi=\Tilde{C}/T$, with $\Tilde{C}=(1-E_F/\Delta)\hat{C}$ the modified Curie constant.
At low temperatures, the susceptibility is temperature independent as Pauli susceptibility, and the asymptote of the finite bandwidth model is $\chi(T) = k_B \hat{C}/E_F$, where $\hat{C}$ is the Curie constant \ref{eq:Curie_constant_FB}.

\section{Susceptibility in the framework of the finite bandwidth model}

\begin{figure}
    \centering
    \includegraphics{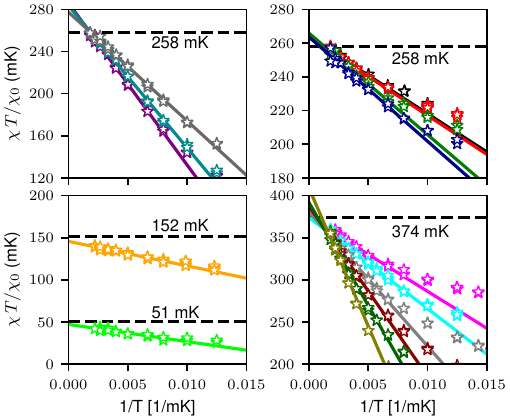}
    \includegraphics{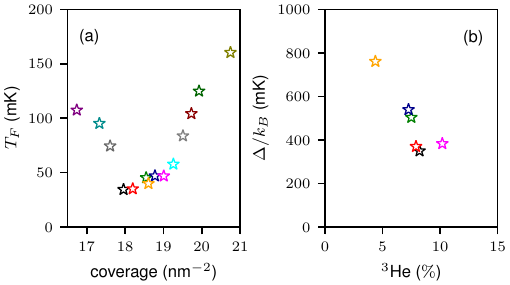}
    \centering\includegraphics{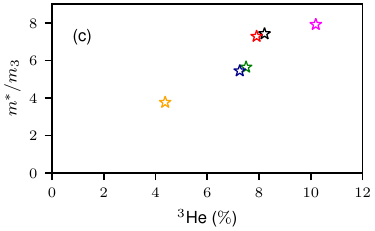}
    \centering\includegraphics{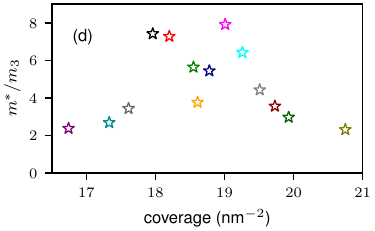}
    \caption{Top panel: correction to the Curie law at the highest temperatures.
    Bottom panel: (a) Fermi temperature $T_F$ as a function of coverage,
    (b) The impuriton bandwidth as a function of \textsuperscript{3}He concentration calculated form $T_F$ according to eq.~\ref{eq:impuritons_Fermi}.
    (c) The band mass according to eq.~\ref{eq:Band_mass} of selected samples as a function of \textsuperscript{3}He concentration and (d) as a function of total coverage.
    } 
    \label{fig:Free_Fermi_gas_asymptote}
\end{figure}

We restrict our analysis of the susceptibility data to the high temperature limit, as shown in Fig.~\ref{fig:Free_Fermi_gas_asymptote}.
In 2D, the free Fermi gas susceptibility is $\chi_0 = \hat{C}/T_F^0 = \mu_0\mu^2mA/\pi\hbar$, and the free Fermi temperature $T_F^0 = \pi\hbar^2N/k_BmA$, making the susceptibility only dependent on area.
$T_F^0 = 505.4\,$mK\,nm\textsuperscript{-2}$\cdot x_3$ is only dependent on \textsuperscript{3}He concentration, and comes out 51, 152, 258 and 374\,mK for the 0.1, 0.3, 0.51, and 0.74\,nm\textsuperscript{-2} coverages, respectively.
In the finite bandwidth model in the r\a'egime where $T>T_F$ and $T<\Delta$, the leading correction to the Curie law of the free Fermi gas at high temperatures is linear
\begin{equation}
    \chi_\infty(T) = \frac{\hat{C}}{T}\left( 1-\frac{T_F}{2T} \right),
\end{equation}
where $T_F=E_F/k_B$ is the Fermi temperature of the \textsuperscript{3}He impuriton gas, determine by number density and bandmass.
In the $\chi T/\chi_0$ \textit{vs.} $T$ coordinates we get
\begin{equation}
    \frac{\chi T}{\chi_0} = T_F^0\left( 1-\frac{T_F}{2T} + ... \right).
\label{eq:free_fermi_gas_model}
\end{equation}
Indeed, the leading correction of our data is linear, Fig.~\ref{fig:Free_Fermi_gas_asymptote}. 
The data should asymptotically approach $T_F^0$ as $1/T\rightarrow 0$, which calculated for the known \textsuperscript{3}He coverages is shown as horizontal lines.
This is indeed the case and validates our ${\chi_0}$ calibration.

% https://robkerr.ai/matplotlib-color-keys/

\definecolor{mypurple}{HTML}{800080}
\definecolor{mydarkcyan}{HTML}{008b8b}
\definecolor{mydimgray}{HTML}{696969}

\definecolor{myred}{HTML}{ff0000}
\definecolor{mygreen}{HTML}{008000}
\definecolor{mydarkblue}{HTML}{00008b}
\definecolor{mymagenta}{HTML}{ff00ff}
\definecolor{mycyan}{HTML}{00ffff}
\definecolor{mygray}{HTML}{808080}
\definecolor{mydarkred}{HTML}{8b0000}
\definecolor{mydarkgreen}{HTML}{006400}
\definecolor{myolive}{HTML}{808000}
\definecolor{mylime}{HTML}{00ff00}
\definecolor{myorange}{HTML}{ffa500}

\begin{table}[!b]
    \centering
    \begin{tabular}{|c|c|c|c|c|}
    \hline
        Symbol & $n_{4}$ [nm$^{-2}$] & $n_3$ [nm$^{-2}$] & $n_{tot}$ [nm$^{-2}$] & $x_3$ [\%] \\ \hline
        \huge$\color{mypurple}\smallstar$      & 16.23 & 0.51 & 16.74 & 10.1\\
        \huge$\color{mydarkcyan}\smallstar$        & 16.82 & 0.51 & 17.33 &  9.1\\
        \huge$\color{mydimgray}\smallstar$    & 17.10 & 0.51 & 17.61 &  8.7\\
        \huge$\color{black}\smallstar$       & 17.45 & 0.51 & 17.96 &  8.2\\
        \huge$\color{myred}\smallstar$         & 17.69 & 0.51 & 18.20 &  8.0\\
        \huge$\color{mygreen}\smallstar$ & 18.04 & 0.51 & 18.55 &  7.6\\
        \huge$\color{mydarkblue}\smallstar$        & 18.27 & 0.51 & 18.78 &  7.4\\
        \huge$\color{mymagenta}\smallstar$     & 18.27 & 0.74 & 19.01 & 10.4\\
        \huge$\color{mycyan}\smallstar$        & 18.52 & 0.74 & 19.26 & 10.0\\
        \huge$\color{mygray}\smallstar$        & 18.77 & 0.74 & 19.51 & 9.7\\
        \huge$\color{mydarkred}\smallstar$      & 18.99 & 0.74 & 19.73 & 9.4\\
        \huge$\color{mydarkgreen}\smallstar$  & 19.19 & 0.74 & 19.93 & 9.2\\
        \huge$\color{myolive}\smallstar$       & 20.01 & 0.74 & 20.75 & n/a\\
        \huge$\color{mylime}\smallstar$        & 18.65 & 0.10 & 18.75 & 1.4\\
        \huge$\color{myorange}\smallstar$      & 18.31 & 0.30 & 18.61 &  4.5\\ \hline
    \end{tabular}
    \caption{Summary of the samples measured by NMR.
    The last column gives \textsuperscript{3}He concentration in the second layer.
    }
    \label{tab:samples}
\end{table}

The Fermi temperature is extracted from the linear fits in Fig.~\ref{fig:Free_Fermi_gas_asymptote} using eq.~\ref{eq:free_fermi_gas_model}, and is shown as a function of total coverage in Fig.~\ref{fig:Free_Fermi_gas_asymptote}\,(a).
The fitting interval is $T>143\,$mK ($1/T<0.007\,$mK$^{-1}$). 
The inferred bandwidth is shown in Fig.~\ref{fig:Free_Fermi_gas_asymptote}\,(b) as a function of the \textsuperscript{3}He concentration for the samples closest to the commensurate coverages.
Clearly these results indicate the breakdown of the independent impuriton model.
From the observed bandwidth of our samples, 400--800\,mK, we get effective mass according to eq.~\ref{eq:Band_mass} to be 4--8\,$m_3$ (where $m_3$ is the bare \textsuperscript{3}He mass) and increasing with concentration.

\begin{figure}
    \centering
    \includegraphics{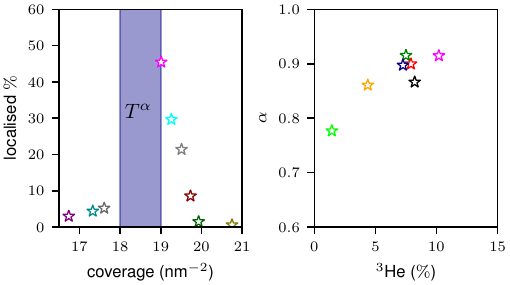}
    \caption{
    (a) The percentage of \textsuperscript{3}He which is localised from the seeming intercept on the $\chi T/\chi_0$ \textit{vs.} $T$ coordinates from Fig~2 in the main body.
    Vertical band shows where no intercept is found and the susceptibility obeys a power law dependence with $\alpha<1$ instead.
    (b) \textsuperscript{3}He concentration dependence of the power $\alpha$ from fits in Fig~2 in the main body.}
    \label{fig:Power_alpha_and_solid_perc}
\end{figure}

\section{Heat capacity in the framework of the finite bandwidth model}

\begin{figure}[!t]
    \centering
    \includegraphics[width=0.8\linewidth]{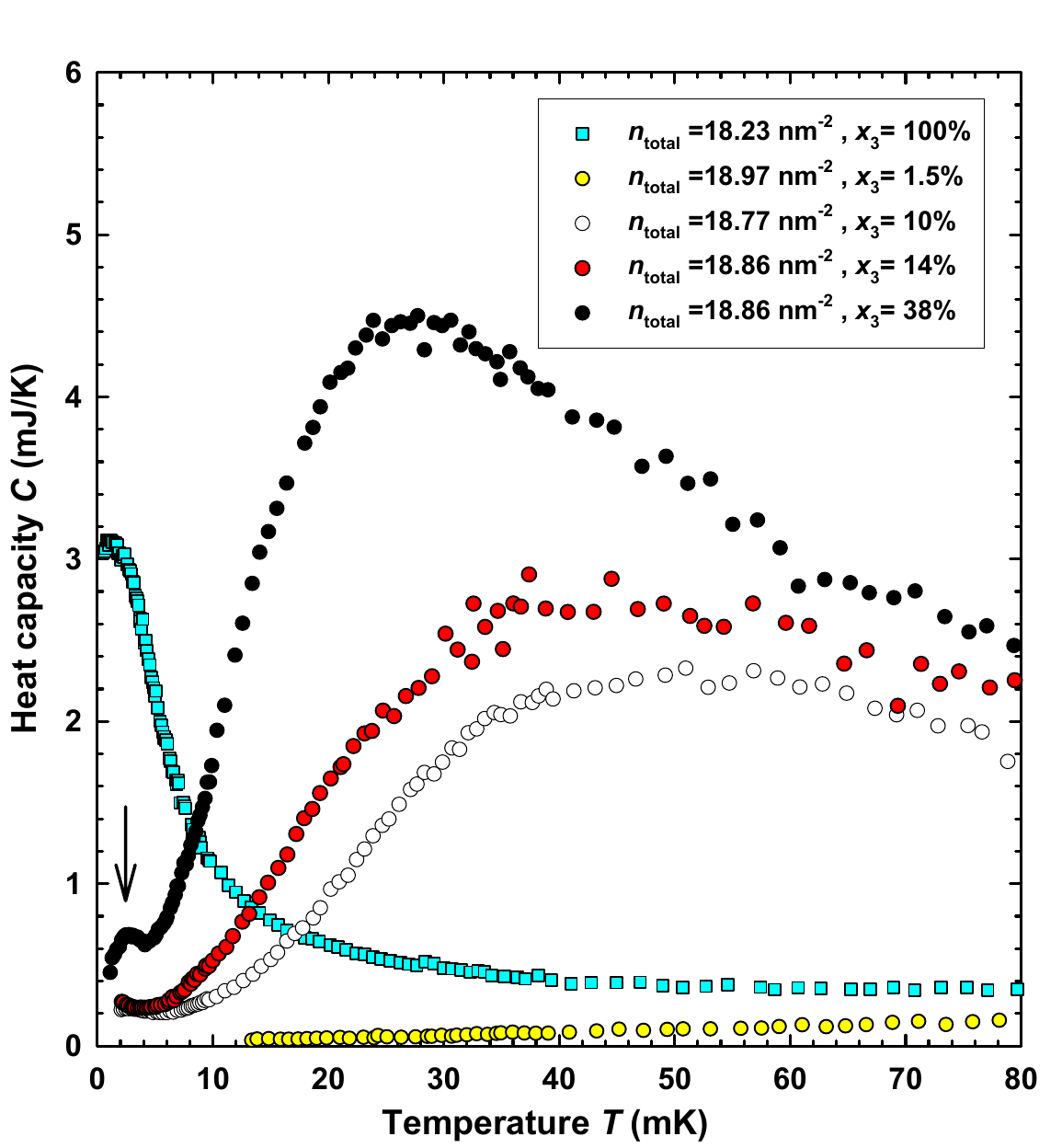}
    \caption{Heat capacity measurements from the vicinity of the commensurate coverages with varying concentration of \textsuperscript{3}He.
    The arrow indicates the ``exchange peak''.
    }
    \label{fig:He_heat_capacity_concentration}
\end{figure}

The temperature dependence of heat capacity shows a broad maximum in the solid region.
A maximum is also predicted by the model, however the position of the peak behaves in the opposite way than predicted.
The measurements shown in Fig.~3 of the main body are with 0.7\,nm$^{-2}$ of \textsuperscript{3}He.
In comparison, Fig.~\ref{fig:He_heat_capacity_concentration} shows the temperature dependence of heat capacity with varying concentration of \textsuperscript{3}He from the vicinity of the commensurate phases.
We observe that the peak goes down in temperature, rather than up as the model which neglects impuriton interactions predicts.
This signals the importance of impuriton interactions, which as discussed are most clearly evidenced by the susceptibility data.
Let us therefore analyze the maximum at the lowest \textsuperscript{3}He concentration, which occurs at the highest temperature of 51\,mK, in the framework of the model.
In this sample, which is 18.07 + 0.70\,nm$^{-2}$, the concentration of \textsuperscript{3}He is $x_3=0.103$.
Locating the correct $E_F/\Delta$ in Fig.~\ref{fig:Finite_band_model_peak_pos_mag} gives $k_BT_p/\Delta=0.104$.
Considering the (broad) peak maximum to be about 50\,mK, we get $k_B\Delta=467\,$mK and $T_F=69\,$mK.
Both these numbers are consistent with the susceptibility result in Fig~\ref{fig:Free_Fermi_gas_asymptote}.

The observed heat capacity shows no evidence of a linear-in-$T$ behaviour characteristic of a fully degenerate Fermi gas.
While the correlated motion of impuritons in bulk crystals is well understood~\cite{Andreev1982}, in that case the ``impuriton gas'' is in the classical r\a'egime.
There is no theoretical treatment of the effect of turning on long range strain interactions in the limit that the impuriton gas is degenerate.
This might be modeled by a reduction in the effective density of states at the Fermi energy, either by introducing a pseudogap, or (by analogy with the ``Coulomb gap'') a density of states which vanishes as a power law in $(E-E_{F})$.
These lead to an activated heat capacity and a power law temperature dependence respectively.
We are unable to discriminate between these possibilities using the present data, which however suggest a characteristic temperature scale of the interaction $\sim$\,10\,mK and not strongly dependent on isotopic concentration.

\section{Isotopic phase separation}

We have argued that the large \textsuperscript{3}He bandwidth in 2D inhibits isotopic phase separation at the \textsuperscript{3}He concentrations we have studied.
This was further explored experimentally by going to significantly higher concentrations.
Partial isotopic phase separation was observed at a much higher \textsuperscript{3}He dose 2.60\,nm$^{-2}$, where at the total helium coverage close to the 7/12 superlattice, a low temperature peak around few mK was observed and attributed to clusters of pure 2D solid \textsuperscript{3}He giving rise to a low temperature magnetic exchange peak.
However, the magnitude of this peak, in comparison with prior heat capacity measurements of a pure \textsuperscript{3}He second layer on a \textsuperscript{4}He first layer~\cite{Ziouzia2004}, showed that only about 50\% of the \textsuperscript{3}He sample was phase separated.
This co-exists with a saturated homogeneous \textsuperscript{3}He-\textsuperscript{4}He solid solution, with estimated concentration of 20\%.
This result is consistent with the absence of phase separation at the lower concentrations we have studied.
We note that phase separation at much higher \textsuperscript{3}He isotopic concentrations was studied by~\cite{Collin2006}.
A previous study of the magnetization of submonolayer isotopic mixture films~\cite{Saunders1992} found clear evidence of isotopic phase separation in the incommensurate solid phase of the film, as well as indirect evidence for delocalization of \textsuperscript{3}He atoms.
This difference in behavior is attributable to a smaller tunneling bandwidth and stronger \textsuperscript{3}He-\textsuperscript{3}He interactions in that case.

% bare mass of 3He atom: 3.016u, u = 1.66e-27 kg -> 5.01e-27 kg

\section{Sample preparation}

The sample preparation is the same in all experimental cells.
The principle is that there are a series of ``thermal annealing'' steps, after every change of total coverage of the helium film on the Grafoil substrate, in order to guarantee an equilibrium state.
This annealing occurs through transport through the vapor.
The vapor pressure above the film is a function of temperature and coverage.
It is established that a vapor pressure of 1\,mbar is effective for this annealing.

On the first cooldown, the cell is first held at between 18.5\,K and 16\,K and a fraction of a \textsuperscript{4}He monolayer added (between 5.6\,nm$^{-2}$ and 6.6\,nm$^{-2}$).
This is the cumulative effect of in shots of defined amount from a room temperature manifold.
This step populates the deep sites in a controlled way.
The sample is then cooled to 10.3\,K and the coverage is increased to 10.3\,nm$^{-2}$, somewhat less than a ``complete'' first layer.
We then perform a \textsuperscript{4}He isotherm at 4.2\,K, and identify point-B to fix our coverage scale (see manuscript).

The sample is then cooled to dilution refrigerator temperature.
All subsequent helium (either \textsuperscript{4}He or \textsuperscript{3}He) is first added at dilution refrigerator temperatures, where the vapor pressure is negligible.
Then the sample is warmed, while monitoring the pressure, until the vapor pressure is around 1\,mbar.
The sample is annealed for several hours.
We cool to take data.
At all the temperatures reported the helium vapor pressure is negligible.
We emphasize that following these procedures there is no \textsuperscript{3}He in the first layer; \textsuperscript{4}He is preferentially bound to the graphite surface because of its higher mass, lower zero point energy, and hence higher binding energy~\cite{Saunders2020}.
It should also be noted that we can only progressively increase the total coverage by adding \textsuperscript{4}He or \textsuperscript{3}He.
We cannot selectively remove one isotope, or reduce total coverage, while maintaining isotopic composition, in a controlled way.

%% uncomment the next two lines to use BiBTeX
%\bibliography{Solidmix13}
%\end{document}


\end{document}